\documentclass[aps,prl,twocolumn,superscriptaddress,showpacs,ams]{revtex4}

\usepackage{graphicx}% complex graphics
\usepackage{bm}      % bold math
\usepackage{amssymb} % math symbols

\newcommand{\imp}{\text{imp}}

\begin{document}

\title{Zero-field Kondo splitting and quantum-critical transition in double quantum dots}

\author{Luis G.\ G.\ V.\ Dias da Silva}
\email{dias@phy.ohiou.edu} \affiliation{Department of Physics and
Astronomy, Nanoscale and Quantum Phenomena Institute, Ohio
University, Athens, Ohio 45701--2979}

\author{Nancy P.\ Sandler}
\affiliation{Department of Physics and Astronomy, Nanoscale and
Quantum Phenomena Institute, Ohio University, Athens, Ohio
45701--2979}

\author{Kevin Ingersent}
\affiliation{Department of Physics, University of Florida,
P.O.\ Box 118440, Gainesville, Florida, 32611--8440}

\author{Sergio E.\ Ulloa}
\affiliation{Department of Physics and Astronomy, Nanoscale and
Quantum Phenomena Institute, Ohio University, Athens, Ohio
45701--2979}

\date{\today}

\begin{abstract}
Double quantum dots offer unique possibilities for the study of
many-body correlations. A system containing one Kondo dot and one
effectively noninteracting dot maps onto a single-impurity
Anderson model with a structured (nonconstant) density of states.
Numerical renormalization-group calculations show that while band
filtering through the resonant dot splits the Kondo resonance, the
singlet ground state is robust. The system can also be
continuously tuned to create a pseudogapped density of states and
access a quantum critical point separating Kondo and non-Kondo
phases.
\end{abstract}


%%72.15.Qm - Scattering mechanisms and Kondo effect
%%73.63.Kv - Electronic transport in nanoscale materials and structures: Quantum Dots
%%73.23.-b - Electronic transport in mesoscopic systems

\maketitle

The Kondo effect, arising from antiferromagnetic correlations
between an unpaired spin and an electron bath \cite{HewsonBook},
can be strongly modified by structure in the host density of
states (DoS). Geometric confinement \cite{KondoBox} and narrow
bands \cite{Hofstetter:R12732:1999} can dramatically change the
Kondo state, with important observable consequences
\cite{Aligia:S1095:2005}. In pseudogapped hosts, where the DoS
vanishes as a power-law at the Fermi energy, a quantum critical
point (QCP) separates the Kondo phase from one at smaller
couplings in which the Kondo effect is completely suppressed
\cite{PseudogapKondo,PseudogapAndersonNRG,Fritz:214427:2004}.

Semiconductor quantum dots provide many opportunities for
systematic investigation of strong-correlation effects
\cite{Mesoreview}. Single quantum dots have allowed controlled
realizations of the Kondo regime of the Anderson impurity problem
\cite{KondoQDExp}. Recent attention has focused on the fascinating
physics promised by double quantum-dot (DQD) systems
\cite{DQDtheory}. For example, DQD experiments have investigated
the effect of interdot ``hybridization'' on Kondo physics
\cite{AChangExpt}, and have beautifully demonstrated the
competition between the Kondo effect and the
Ruderman-Kittel-Kasuya-Yosida interaction among localized spins
\cite{Craig:565}. DQD setups have also been proposed to realize
the unusual non-Fermi-liquid properties associated with the
two-channel Kondo effect \cite{QD2chKondo}.

In this paper, we propose DQDs as a versatile experimental
realization of an impurity coupled to an electron bath having a
structured (nonconstant) DoS. Devices with one dot (``dot 1'') in
the Kondo regime and the other (``dot 2'') close to resonance with
the leads can be designed to produce an effective DoS having sharp
resonances and/or pseudogaps near the Fermi energy. These features
are shown to strongly modify the Kondo state, resulting in a wide
range of DQD behavior, which we explore using numerical
renormalization-group methods.

We show that when dot 1 is coupled to the leads \textit{only
through} dot 2, the Kondo resonance on dot 1 develops a sizable
splitting. Unlike magnetic fields, which produce similar
splittings, the ``band filtering'' introduced by the connecting
dot preserves the Kondo singlet ground state, and results in a
finite Kondo temperature for complete screening of the magnetic
moment on dot 1.

A second configuration, involving coherent dot-dot coupling via
the leads, can mimic an Anderson impurity in a pseudogapped host.
The device can be tuned by varying gate voltages to a QCP
separating Kondo-screened and free-local-moment ground states.
Such DQDs offer an attractive experimental setting for systematic
study of boundary quantum phase transitions.

\begin{figure}[t]
\includegraphics[totalheight=0.4\columnwidth,width=0.6\columnwidth]{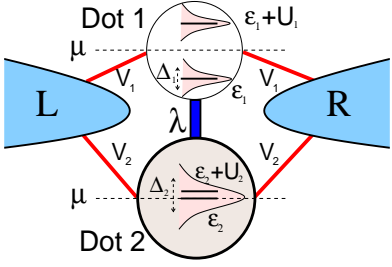}
\caption{\label{fig:DDot}
(color online) Schematic of the DQD system. Dot 1 is Kondo-like
($-\varepsilon_1, \varepsilon_1+U_1 \gg \Delta_1 =\pi\rho_0
V_1^2$, where $\rho_0$ is the lead DoS), while dot 2 can be
treated as a single, noninteracting ($U_2\ll \Delta_2 =\pi\rho_0
V_2^2$) resonant level.}
\end{figure}

The DQD consists of dots 1 and 2 connected to left ($L$) and right
($R$) leads as well as to each other, as shown schematically in
Fig.\ \ref{fig:DDot}. We focus on situations in which dot 1 is
tuned to have an odd number of electrons in a Coulomb blockade
valley, so that it has an unpaired spin, while near-resonant
transport through dot 2 is dominated by a single level
\cite{Dot2Note} having a dot-lead coupling greater than the
charging energy, so that the dot can be considered noninteracting.
Our Hamiltonian is therefore
\begin{eqnarray}
\label{Eq:Hamiltonian}
  H & = &
    \sum_{i,\sigma} \varepsilon_i n_{i\sigma} + U_1 n_{1 \uparrow}n_{1 \downarrow}
    + \sum_{\sigma} \! \left( \lambda a^{\dagger}_{1 \sigma}
    a_{2\sigma} + \mbox{H.c.} \! \right) \nonumber \\
  & + & \sum_{j,{\bf k},\sigma} \varepsilon_{\bf k} c^{\dagger}_{j {\bf k} \sigma}
    c_{j {\bf k} \sigma} + \sum_{i,j,{\bf k},\sigma} \!\! \left(
    V_{ij} \, a^{\dagger}_{i \sigma} c_{j {\bf k} \sigma} + \mbox{H.c.}
    \! \right) \! ,
\end{eqnarray}
where $a^{\dagger}_{i\sigma}$ creates a spin-$\sigma$ electron in
dot $i$ ($=1,2$), $n_{i\sigma}=a^{\dagger}_{i \sigma}a_{i\sigma}$,
and $c^{\dagger}_{j {\bf k} \sigma}$ creates a spin-$\sigma$
electron of wavevector ${\bf k}$ and energy $\varepsilon_{\bf k}$
in lead $j$ ($=L,R$). For simplicity, we take the dot-lead
couplings $V_{ij}$ to be ${\bf k}$-independent. We further assume
$V_{iR} = V_{iL}$, in which case the dots couple to the leads only
in the symmetric combination $c_{{\bf k}\sigma} = (c_{L{\bf
k}\sigma} + c_{R{\bf k}\sigma})/\sqrt{2}$ and Eq.\
(\ref{Eq:Hamiltonian}) describes double dots effectively coupled
to a \textit{single} lead, with $V_i\equiv\sqrt{2}V_{iL}$.
Near-symmetric couplings can be achieved experimentally by
appropriate tuning of the dot-lead tunneling gate voltages (see,
e.g., \cite{AChangExpt}). $V_i$ and the dot-dot coupling $\lambda$
are taken to be real and positive.

The Green's function (GF) for dot 1 is
$G_{11}(\omega)\equiv\langle\langle a_{1 \sigma}\!:\!
a^{\dagger}_{1\sigma} \rangle \rangle =
(1+U_1\Gamma_{11(\omega)})G_{11}^{(0)}(\omega)$, where $\omega$ is
the energy relative to the common chemical potential $\mu=0$ in
the leads, $\Gamma_{11}(\omega)=\langle\langle n_{1,-\sigma} a_{1
\sigma} \!\! : \! a^{\dagger}_{1 \sigma}\rangle\rangle$, and
$G_{11}^{(0)}(\omega)$ is the noninteracting GF for dot 1 in the
presence of dot 2:
\begin{eqnarray}
\label{Eq:non-intGF}
  \left[ G_{11}^{(0)}(\omega) \right]^{-1} = \left[
    G_1^{(0)}(\omega) \right]^{-1} - G_2^{(0)}(\omega) \times
    \qquad \qquad \mbox{} \nonumber\\
  \left[ \lambda^2 + \lambda
    \sum_{{\bf k}} \frac{2V_1V_2}{\left(\omega-\varepsilon_{{\bf k}}\right)}
    + \sum_{{\bf k},{\bf k^\prime}}
    \frac{V_1^2 V_2^2}{\left(\omega-\varepsilon_{{\bf k}}\right)
    \left(\omega-\varepsilon_{{\bf k^\prime}}\right)} \right] ,
\end{eqnarray}
$G^{(0)}_{i}(\omega) = \left[
\omega\!-\!\varepsilon_i\!-\!\sum_{\bf k}
V_i^2/(\omega\!-\!\varepsilon_{\bf k})\right]^{-1}$ being the
noninteracting GF for dot $i$ in the absence of the other dot.

Hereafter, we assume a constant DoS $\rho_0$ in the leads. In the
wide-band limit (half bandwidth $D\gg|\omega|$), we can formally
write $\left[ G_{11}(\omega) \right]^{-1} = \omega-\varepsilon_1 -
\Sigma_{11}^*(\omega) + \Lambda(\omega) + i\Delta(\omega)$, where
$\Sigma_{11}^*(\omega)$ is the proper self-energy,
$\Lambda(\omega) = \pi \rho_2(\omega) \left[ (\lambda^2-
\Delta_1\Delta_2) (\omega-\varepsilon_2)/\Delta_2 - 2 \lambda
\sqrt{\Delta_1 \Delta_2} \right]$, and
\begin{equation}
\label{Eq:Delta}
  \Delta(\omega) = \pi \rho_2(\omega) \left[\lambda + (\omega -
    \varepsilon_2)
    \sqrt{\Delta_1/\Delta_2}
    \right]^2,
\end{equation}
with $\Delta_i = \pi \rho_0 V^2_i$ and
$\rho_2(\omega)=\Delta_2/\{\pi[(\omega-\varepsilon_2)^2 +
\Delta_2^2]\}$.

All information on the coupling of dot 1 to the leads and to dot 2
enters $G_{11}$ through $\Lambda(\omega)$ (which essentially
renormalizes the single-particle energy $\varepsilon_1$) and, more
importantly, through $\Delta(\omega)$. This provides a mapping of
the DQD onto a single Anderson impurity coupled to a Fermi system
with an {\em effective hybridization function} $\Delta(\omega)$,
which ``filters'' the band states seen by dot 1 and modifies its
coupling to the leads. We have solved this interacting model using
an extension of the numerical renormalization-group (NRG) method
\cite{KrishnamurthyWW80_1} designed to handle arbitrary conduction
band shapes \cite{PseudogapAndersonNRG}.

In order to understand the effects of the \textit{nonconstant}
effective hybridization $\Delta(\omega)$, we consider (i) a
\textit{side-dot} configuration, in which dot 1 is coupled to the
leads only through dot 2 ($\lambda \ne 0$, $V_1=0$); (ii) a
\textit{parallel} configuration, in which interdot interactions
take place only indirectly via the leads ($\lambda=0$, $V_1\ne
0$); and (iii) a more general \textit{fully connected}
configuration ($\lambda\ne 0$, $V_1\ne 0$).

\begin{figure}[t]
\includegraphics*[height=0.85\columnwidth,width=1.0\columnwidth]{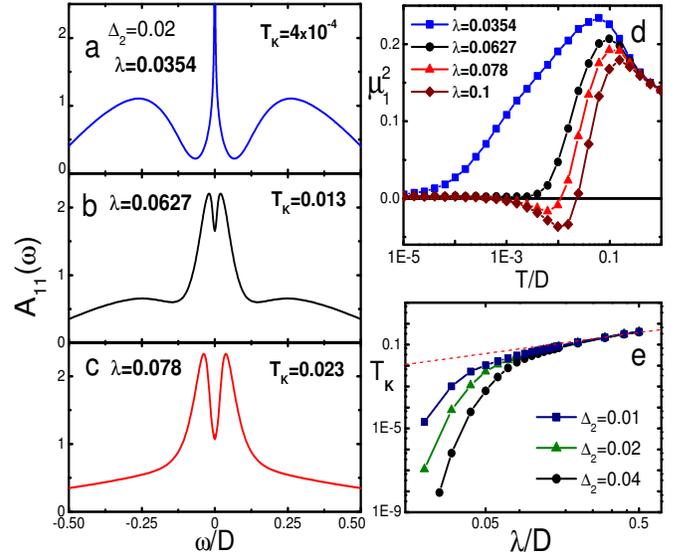}
\caption{(color online)
Side-dot configuration. (a)--(c) \mbox{Dot-1} spectral density for
$U_1\!=\!-2\varepsilon_1\!=\!0.5$, $\varepsilon_2\!=\!0$, and
$\Delta_2\!=\!0.02$. Splitting appears in $A_{11}(\omega)$ for
larger $\lambda$ such that $T_K \agt \Delta_2/\sqrt{2}$. (d)
Effective moment $\mu_1^2$ vs $T$. (e) $T_K$ vs $\lambda$. For
large $\lambda$, $T_K\propto\lambda$ (dashed). (Energies and $T_K$
in units of $D$.)}
\label{fig:LorentG002}
\end{figure}
\noindent (i) In the \textit{side-dot configuration}, the
effective hybridization $\Delta(\omega)=\pi \rho_2(\omega)
\lambda^2$, so the system maps onto an Anderson impurity coupled
to a Lorentzian DoS. The case $\varepsilon_2=0$, which places the
peak in $\Delta(\omega)$ at the Fermi energy $\omega=0$, yields
the most striking properties.

Figure \ref{fig:LorentG002} presents results for $\varepsilon_2=0$
and $U_1=-2\varepsilon_1$, for which parameters the model exhibits
strict particle-hole ($p$-$h$) symmetry. Figs.\
\ref{fig:LorentG002}(a)--(c) show the spectral density
$A_{11}(\omega)=-\mbox{Im} \, G_{11}(\omega)/\pi$ for
$\Delta_2=0.02D$ and different interdot couplings $\lambda$. For
small $\lambda$ [Fig.\ \ref{fig:LorentG002}(a)], the spectral
density resembles that for a constant DoS, its main features being
broad Hubbard bands centered near $\omega=\varepsilon_1$ and
$\varepsilon_1+U_1$, and a sharp resonance at $\omega=0$ having a
width of order the Kondo temperature $T_K$ (defined below). In
this regime, the nonconstant $\Delta(\omega)$ manifests itself
through a generalized Fermi-liquid relation \cite{Dias:inPrep:}
$A_{11}(0)=\cos^2\varphi/[\pi\Delta(0)]$, where $\varphi =
\int_{-\infty}^0 \left[ (d\Delta/d\omega)\mathrm{Re}\,
G_{11}(\omega)-(d\Lambda/d\omega)\mathrm{Im}\, G_{11}(\omega)
\right] d\omega$, i.e., $A_{11}(0)$ is smaller by a factor of
$\cos^2\varphi$ than the standard result \cite{HewsonBook} for a
flat band with the same $\Delta(0)$.

For larger $\lambda$, such that $T_K \agt \Delta_2/\sqrt{2}$, the
spectral density is qualitatively different. The Kondo resonance
initially rises under the influence of the relatively weak
hybridization found for $|\omega| \agt T_K$. However, the upturn
in $\Delta(\omega)$ at $|\omega|\alt \Delta_2$ causes
$A_{11}(\omega)$ to drop to satisfy $A_{11}(0) \le
1/[\pi\Delta(0)]$ (see above), splitting the resonance into two
distinct peaks [Fig.\ \ref{fig:LorentG002}(b)]. As $\lambda$
increases further, the dip deepens and the Kondo peaks move out,
eventually subsuming the Hubbard bands [Fig.\
\ref{fig:LorentG002}(c)].

The splitting of the Kondo peak and the suppression of $A_{11}(0)$ might be
supposed to signal the destruction of the Kondo singlet ground state (as is
the case in a magnetic field).
However, this interpretation is refuted by the NRG many-body spectra, which
show that the Kondo ground state is reached even for $T_K\gg \Delta_2$.

The progressive screening of the localized spin with decreasing
temperature $T$ can be seen Fig.\ \ref{fig:LorentG002}(d), which
plots the square of the effective free moment on dot 1,
$\mu_1^2(T)\equiv T\chi_{\imp}$, where $\chi_{\imp}$ is the dot
contribution to the zero-field susceptibility. The behavior for
$T_K \alt \Delta_2$ follows that for a constant DoS
\cite{KrishnamurthyWW80_1}, crossing at $T\approx|\varepsilon_1|$
from the free-orbital regime ($\mu^2_1 \approx 1/8$) to the
local-moment (LM) regime ($\mu^2_1 \approx 1/4$), and then for $T
\ll T_K$, to the strong-coupling (SC) Kondo limit in which the
magnetic moment is totally screened ($\mu_1=0$).

For $T_K \agt \Delta_2$, the system still reaches the Kondo fixed
point with $\mu_1^2 = 0$ as $T\rightarrow 0$. However, the dot
spin now exhibits an interesting window of \textit{diamagnetic}
behavior ($\chi_{\imp} < 0$), which becomes more pronounced as
$T_K$ increases [Fig.\ \ref{fig:LorentG002}(d)]. This diamagnetic
region arises from a negative term $\propto d\rho_2(\omega)/d
\omega$ in $\chi_{\imp}$ \cite{Dias:inPrep:}.

Since $\mu_1^2$ in all cases passes from 1/8 at high temperatures
to 0 at $T=0$, we define the Kondo temperature using the standard
criterion $\mu_1^2(T_K)=0.0701$ \cite{KrishnamurthyWW80_1}. $T_K$,
shown in Fig.\ \ref{fig:LorentG002}(e) for three different
$\Delta_2$ values, increases rapidly for small $\lambda$, and
satisfies $T_K\propto\lambda$ in the noninteracting narrow-band
limit $\lambda\gg \Delta_2, U_1/2$.

The splitting observed in the Kondo resonance and the diamagnetic
region in $\mu^2_1(T)$ are produced by the sharp peak in
$\rho_2(\omega)$ at $\omega=0$ resulting from the resonance in dot
2. The resonance acts as a \textit{filter} for the higher-energy
states in the leads, reducing the effective conduction bandwidth
connected to dot 1. This interpretation, which is consistent with
similar findings of a negative $\chi_{\imp}$ in narrow-band
systems \cite{Hofstetter:R12732:1999}, clearly differentiates the
side-dot behavior from the peak splittings due to coherent
coupling to a second Anderson impurity and the resulting
suppression of the singlet state \cite{AChangExpt}.

\begin{figure}[t]
\includegraphics*[height=0.84\columnwidth,width=1.0\columnwidth]{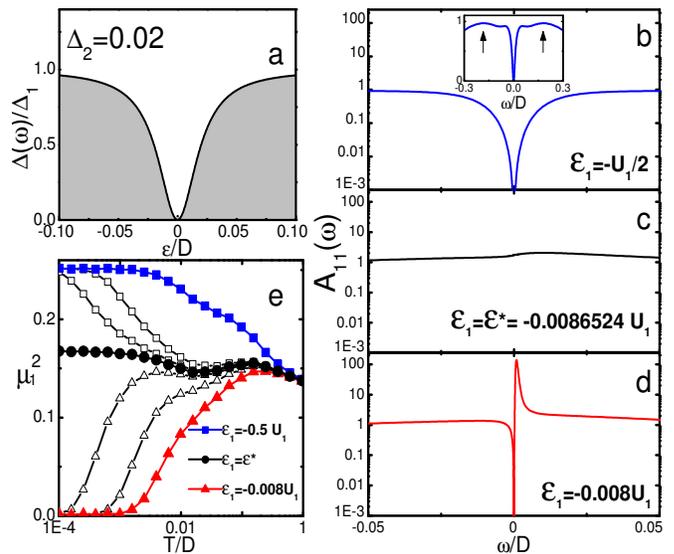}
\caption{(color online) Parallel DQD
with $U_1=0.5$, $\Delta_1=0.05$, $\Delta_2=0.02$, and
$\varepsilon_2 = 0$. (a) Hybridization $\Delta(\omega)$ vanishes
as $|\omega|^2$ at $\omega=0$. (b)--(d) Dot-1 spectral density
$A_{11}(\omega)$, and (e) effective moment $\mu_1^2(T)$, for
various $\varepsilon_1$. For $\varepsilon_1\approx -U_1/2$, no
Kondo effect occurs: $A_{11}(0)=0$ in (b), and $\mu_1^2(0)= 1/4$
[$\scriptscriptstyle\blacksquare$ and $\scriptscriptstyle\square$
in (e)]. This LM phase is separated by a QCP at
$\varepsilon_1=\varepsilon^{*}>-U_1/2$, characterized by a
featureless $A_{11}(\omega)$ in (c) and $\mu_1^2(0)=1/6$
[$\bullet$ in (e)], from the SC phase [(d),
$\scriptstyle\blacktriangle$ and $\scriptstyle\triangle$ in (e)]
in which $A_{11}$ vanishes at $\omega=0$ and peaks at some
$\omega>0$.}
\label{fig:PDipQCPcols}
\end{figure}

The side-dot behavior can also be understood as arising from
\textit{interference} between resonances. This can be seen by
considering the noninteracting spectral density
$A^{(0)}_{11}=-\mbox{Im} \, G^{(0)}_{11}/\pi$ for
$\varepsilon_1=\varepsilon_2$. For $\lambda<\Delta_2/\sqrt{2}$,
$A^{(0)}_{11}$ has a single peak (width $\sim\lambda^2$) at
$\omega=\varepsilon_2$; whereas for $\lambda>\Delta_2/\sqrt{2}$,
there are two peaks at
$\omega=\varepsilon_2\pm\sqrt{\lambda^2-\Delta^2_2/2}$, arising
from interference between the $\omega=0$ single-particle
resonances on the two dots. The NRG results for the interacting
case are closely analogous: For $T_K \alt \Delta_2/\sqrt{2}$,
$A_{11}$ has a single peak (width $\sim T_K$) at
$\omega=\varepsilon_2=0$, while for $T_K \agt
\Delta_{2}/\sqrt{2}$, there are two peaks, in this case resulting
from interference between the $\omega=0$ many-body Kondo resonance
and the $\omega=0$ single-particle resonance in $\rho_2$. The
separation of the peaks in $A_{11}$ increases with $T_K -
\Delta_{2}$, further supporting the analogy.

\noindent (ii) The \textit{parallel configuration} ($\lambda=0$)
exhibits very different behavior. For $\varepsilon_2=0$, the
hybridization $\Delta(\omega)$ [Eq.\ (\ref{Eq:Delta})]
\textit{vanishes} at $\omega=0$ as $|\omega|^r$ with $r=2$ [see
Fig.\ \ref{fig:PDipQCPcols}(a)], and the DQD setup maps onto an
Anderson impurity in a pseudogapped host \cite{V1RV2LeqV1LV2R}.
The properties of such system depend strongly on the exponent $r$
and on the presence or absence of $p$-$h$ symmetry
\cite{PseudogapAndersonNRG,Fritz:214427:2004}. For $r=2$, the SC
phase is inaccessible at $p$-$h$ symmetry, but away from this
special limit, a QCP separates SC and LM phases.

In the $p$-$h$-symmetric case $\varepsilon_1 = -U_1/2$, the Kondo
resonance in $A_{11}$ disappears completely [Fig.\
\ref{fig:PDipQCPcols}(b)], but Hubbard bands are still present
(arrows in inset). $A_{11}$ vanishes at $\omega=0$ as $\omega^2$,
and $\mu_1^2(0) = 1/4$ [Fig.\ \ref{fig:PDipQCPcols}(e)], as
expected in the LM phase \cite{PseudogapAndersonNRG} where no
Kondo effect occurs. When $p$-$h$ symmetry is broken by increasing
$\varepsilon_1$, the same qualitative behavior persists [squares
in Fig.\ \ref{fig:PDipQCPcols}(e)] until the QCP is reached at
$\varepsilon_1=\varepsilon^*$, where $A_{11}$ is nearly
featureless [Fig.\ \ref{fig:PDipQCPcols}(c)] and $\mu_1^2(0)=1/6$
[circles in Fig.\ \ref{fig:PDipQCPcols}(e)]. For
$\varepsilon_1>\varepsilon^*$, the system enters the SC phase in
which $\mu_1^2(0)=0$ [triangles in Fig.\
\ref{fig:PDipQCPcols}(e)], indicating complete Kondo screening of
dot 1, and $A_{11}$ again goes to zero as $\omega^2$ at
$\omega=0$, but (in contrast to the LM phase), there is a distinct
peak at positive $\omega$ [Fig.\ \ref{fig:PDipQCPcols}(d)].

\noindent (iii) In the \textit{fully connected configuration}
($V_1$, $V_2$, and $\lambda$ all nonzero), $\Delta(\omega)$ has an
asymmetric Fano-like shape, peaking at $\varepsilon_2 \! + \!
\sqrt{\Delta_1\Delta_2^3} /\lambda$, and vanishing as
$(\omega\!-\!\omega_0)^2$ at $\omega_0=\varepsilon_2 \!-\!
\lambda\sqrt{\Delta_2/\Delta_1}$. Here, the DQD properties can be
controlled not only by tuning $\lambda$, $\Delta_1$, or $\Delta_2$
[as in (i) and (ii)], but also by using external gate voltages to
vary $\varepsilon_2$.

\begin{figure}[t]
\includegraphics*[height=0.9\columnwidth,width=1.0\columnwidth]{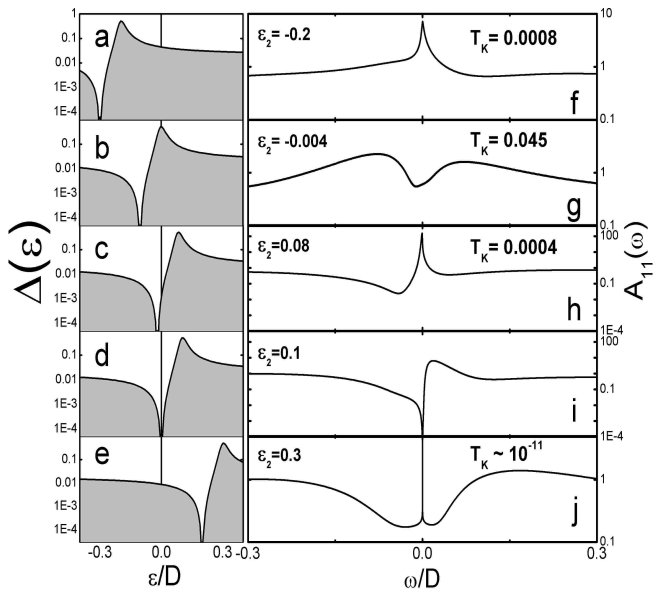}
\caption{
Fully connected DQD with
$U_1=-2\varepsilon_1=0.5$, $\Delta_1=\Delta_2=0.02$, and
$\lambda=0.1$. (a)--(e) Hybridization $\Delta(\omega)$ for various
$\varepsilon_2$, and (f)--(j) the corresponding $A_{11}(\omega)$. An
asymmetric Kondo peak appears when $\Delta(\omega)$ is featureless
near $\omega=0$ (a, e). The peak splits as the resonance in
$\Delta(\omega)$ approaches $\omega=0$ (b), reappears in the
intermediate region (c), and disappears altogether when a
pseudogap forms (d).}
\label{fig:PAllG002g002l01}
\end{figure}

Figures \ref{fig:PAllG002g002l01}(a)--(e) show the evolution of
$\Delta(\omega)$ as $\varepsilon_2$ shifts across the Fermi energy.
Generically, the dot-1 spectral density features a
$p$-$h$-asymmetric Kondo resonance, whose width $T_K$ is primarily
determined by $\Delta(0)$ [Figs.\
\ref{fig:PAllG002g002l01}(f),(h),(j)]. However, when the peak of
$\Delta(\omega)$ approaches $\omega=0$, $T_K$ is enhanced and the
Kondo peak splits [Fig.\ \ref{fig:PAllG002g002l01}(g)]. By
contrast, when the zero of $\Delta(\omega)$ approaches the Fermi
energy, the Kondo temperature decreases, and if $\omega_0$ is
tuned to zero, pseudogap behavior is recovered with complete
suppression of the Kondo peak [Fig.\
\ref{fig:PAllG002g002l01}(i)].

In summary, we have shown that double quantum-dot (DQD) systems
with one of the dots in the Kondo regime can be tailored
experimentally to explore the effects of a nonconstant density of
states (DoS) on the many-body ground-state properties. In setups
where the Kondo dot is decoupled from the reservoirs, the Kondo
resonance on that dot undergoes zero-field splitting for large
interdot coupling. This can be understood as the coherent
interaction between the many-body Kondo state and a
single-particle resonance in the second dot. Although the Kondo
peak in the spectral density at the Fermi energy is suppressed,
the Kondo singlet state is robust, and the localized spin is
completely screened at low temperatures. In this regime, the
system also passes through a temperature window of diamagnetic
behavior, similar to that seen in narrow-band systems.

For weak interdot couplings, the Kondo state is suppressed by the
presence of a pseudogap in the effective DoS, and a quantum phase
transition takes place between local-moment and Kondo-screened
phases. A quantum critical point on the boundary between these
phases can be reached by appropriate tuning of the experimental
parameters. Thus, DQD systems provide a rare example of a
controlled setting in which to investigate quantum critical
behavior in a strongly correlated system.

This work was partially supported by NSF DMR--0312939 (K.I.) and
NSF-IMC/NIRT (L.D.S., N.S., S.U.).


\begin{thebibliography}{10}
\bibitem{V1RV2LeqV1LV2R}
  This mapping holds even for $V_{iR} \ne V_{iL}$ provided that
  $V_{1L} V_{2R} = V_{1R} V_{2L}$; otherwise, $\Delta(0)$
  is small but nonzero.
\end{thebibliography}
\end{document}